\documentstyle[12pt,epsfig]{article}
\begin{document}
\vspace{1.cm}
\begin{center}
\    \par
\    \par
\    \par
\     \par

        {\bf{
HIGH ENERGY PHYSICS INSTITUTE\\ TBILISI I. JAVAKHISHVILI STATE UNIVERSITY }}

\end{center}
\   \par
\    \par
\hspace{9.9cm} {\it{                     Licensed as a Manuscript}}
\   \par
\    \par
\par
\begin{center}
{\large { \bf{L. Kharkhelauri} }}\par
\end{center}
\par
\   \par
\   \par
\    \par
\begin{center}
                 {\large {    \bf{
Study of Multiparticle Azimuthal Correlations in Central
Nucleus-Nucleus Collisions at energy of 3.7 GeV per nucleon}}}
\end{center}
\   \par
\   \par
\    \par
\begin{center}
                     \bf{
01.04.16-Physics of Atoms, Atomic\\
Nuclei and Elementary Particles}
\end{center}
\   \par
\   \par
\    \par
\par
\begin{center}
                     \bf{
A b s t r a c t  $~~$   o f $~~$     D i s s e r t a t i o n \\
For the Fulfilment of Scientific Degree of the Candidate\\
in Physical and Mathematical Sciences}
\end{center}
\ \par
\   \par
\   \par
\    \par
\ \par
\   \par
\   \par
\    \par
\begin{center}
                     \bf{TBILISI,  2001  }
\end{center}
\pagebreak
\begin{center}
{\large { \bf{
The research is fulfilled in High Energy Physics Institute of Tbilisi State
University} }}
\end{center}
\   \par
\   \par
Supervisors:  \hspace{4.9cm} {\bf{  L. CHKHAIDZE}}
\par
\hspace{2.9cm}   {Candidate of Physical and Mathematical Sciences}
\   \par
\   \par
\hspace{7.5cm}               {\bf{  T.  DJOBAVA}}
\par
\hspace{2.9cm}   {Candidate of Physical and Mathematical Sciences}
\   \par
\   \par
Referees:   \hspace{5.7cm}    {\bf{ V. GARSEVANISHVILI}}
\par
\hspace{2.9cm}                Doctor of Physical and Mathematical Sciences, Professor

\hspace{2.9cm}                01.04.16-Physics of Atoms, Atomic Nuclei and Elementary Particles
\   \par
\   \par
\hspace{7.5cm}                {\bf{ M. SVANIDZE}}
\par
\hspace{2.9cm}                Candidate of Physical and Mathematical Sciences
\par
\hspace{2.9cm}                01.04.16-Physics of Atoms, Atomic Nuclei and Elementary Particles
\   \par
\   \par
\   \par
\   \par
Scientific Secretary of the Dissertation Council,
\par
{Doctor of Physical and Mathematical Sciences}
\hspace{2.5cm} {\bf{  T.  KERESELIDZE}}
\pagebreak
\par
\underline{Subject of the Research Work}.
Relativistic nucleus--nucleus collisions are
very well suited for investigation of excited nuclear matter properties under
extreme conditions of high densities and high temperatures, which are the
subject of intense studies both experimentally and theoretically. A new
collective phenomena have been revealed in high energy heavy ion collisions,
which allow to study the nuclear equation of state (EOS) and offers an unique
information about the possible phase transition in the nuclear matter as well.
\par
The experimental discovery of such states is impossible without understanding
the mechanism of collisions and studying the characteristics of multiparticle
production in nucleus-nucleus interactions. During last 10-15 years several
theoretical models have been proposed for the study of nucleus-nucleus
collisions at high energy. It is assumed that, the intranuclear cascade model
(CAScade Intranuclear Model Made in Rossendorf - CASIMIR) and Quark Gluon
String Model (QGSM) allow to investigate hadron-nuclear, hadron-nucleus and
nucleus--nucleus collisions at wide energy interval and especially in the
range of intermediate energy {$\sqrt{s} \leq$ 4 GeV).
\par
In order to study the dynamics of the relativistic nucleus-nucleus collisions
in the series of experiments the inclusive spectra of kinematical variables
have been obtained for secondary protons and mesons. It is interesting to
study the correlation between Lorenz invariant variables, such as the
rapidity $Y$ and transverse momentum $P_{T}$ of the produced particles. Interesting
information about  the collision process and its space-time evolution can be
obtained from the analysis of secondary particles angular distributions with
the aim to determine the degree of anisotropy in particles emission. It is
also important to investigate the energetic spectra of  secondary particles
at different intervals of emission angle $\theta_{lab.}$
\par
In  the energy range of (0.5 - 10) GeV/nucleon the nuclear matter is
compressed and heated more than at lower beam energies. Two different
signatures of collective flow have been predicted by nuclear shock wave
models and ideal fluid dynamics: 1) the bounce-off of the compressed matter
in the reaction plane and 2) the squeeze-out of the participant matter out
of the reaction plane. Collective flow has been already observed for protons,
light nuclei, pions, kaons and ($\Lambda$-hyperons in nucleus--nucleus collisions at
the energies: of (0.1-1.8) GeV/nucleon at BEVALAC (Lawrence Berkeley
Laboratory - USA), GSI/SIS  (Darmstadt - Germany), 3.7 GeV/nucleon at Dubna
(Joint Institute Nuclear Research - Russian), 2 - 14 GeV/nucleon AGS BNL
(Brookhaven National Laboratory - USA) and 60 and 200 GeV/nicleon at CERN SPS
(European Organization for Nuclear Research - Switzerland), and more recently
by the STAR and PHENIX at RHIC BNL at $\sqrt{S_{NN}}$ =130 CeV/nucleon. Many different
methods have been proposed for the study of the flow effects in relativistic
nuclear collisions, of which the most commonly used are the directed
transverse momentum analysis technique developed by P. Danielewicz and
G. Odyniec and the method of the Fourier expansion of azimuthal distributions
proposed by Demoulins et al. and Voloshin and Zang.
\par
The investigation of the collective effects can be also studied in
multiparticle azimuthal correlations, because it gives the information about
space-time evolution of the colliding system. In order to study these
correlations between the protons and between the $\pi^{-}$ mesons, the phase space
can be divided into two hemispheres. The division can be performed by use
of the rapidity and by the emission angle - the forward and backward emitted
particles and by energy - the slow and fast particles as well. The azimuthal
correlations have been already observed between protons and between pions in
nucleon-nucleus and nucleus--nucleus collisions at energies of
(0.4-1.8) GeV/nucleon at BEVALAC, 3.7 GeV/nucleon at Dubna and 60 and
200 GeV/nucleon at CERN SPS.
\par
\underline{Aims of the Thesis}.
\\
$\bullet$
The study of the properties of $\pi^{-}$ mesons produced in Mg-Mg collisions at
energy of E=3.7 GeV/nucl at a SKM-200--GIBS set-up of JINR/DUBNA.
Investigations of the correlation between Lorentz invariant variables: the
rapidity $Y$ and the transverse momentum $P_{T}$. The analysis of angular
distribution of $\pi^{-}$ mesons and the study of the dependence of $\pi^{-}$
mesons emission anisotropy coefficient $a$ on the kinetic energy.
Investigation of emitted $\pi^{-}$ mesons energetic spectra in different
intervals of emission angle $\theta_{lab}$ in the laboratory and centre of
mass systems (c.m.s.).
\\
$\bullet$
Investigation of collective flow effects of $\pi^{-}$ mesons produced in
central C-Ne, Mg-Mg, C-Cu and O-Pb collisions at energy of E=3.7 GeV/nucl
by use of
transverse momentum technique of P. Danielewicz and G. Odyniec. The study of
the dependence of the flow parameter $F$, as the slope at midrapidity, on the
target mass numbers ($A_{T}$).\\
$\bullet$
Investigation of the collective effects by the multiparticle azimuthal
correlations between protons and between pions in central C-Ne, Mg-Mg, C-Cu
and O-Pb collisions. The study of the dependence of the correlation function
on the angles $\Delta\varphi$  between the sumed transverse vectors for particles 
emitted
in the forward and backward hemispheres.\\
$\bullet$
Comparision of the obtained experimental results with predictions of
intranuclear cascade model (CASIMIR) and Quark Gluon String Model (QGSM).
\par
\underline{Scientific Novelty}.
\\
$\bullet$
A detailed study of pion production in central Mg-Mg collisions have been
carried out: a) Correlations between $P_{T}$ and $Y$ have been studied; b) The
anisotropy coefficient $a$ of $\pi^{-}$ mesons has been obtained from the angular
distributions of pions. Its dependence on $E_{kin}/E_{kin}^{max}$ in the c.m.s. has
been studied, c) The dependence of the energy spectra of $\pi^{-}$ mesons
emitted
in an angular interval of $0^{o}\leq \theta_{lab} \leq 180^{o}$ in the
laboratory system and $\theta_{C.M.S.} = 90^{o} \pm 10^{o}$,
$90^{o} \pm 20^{o}$ emission angles in c.m.s. on $A_{P}, A_{T}$ have been
investigated for Mg-Mg, He(Li, C), C-Ne, C-Cu, (C,O)Pb collisions. The values
of fireball model parameters, the temperature $E_{o}^{\perp}$ and velocity
$\beta$ of the
cluster formed during the collisions, which is emitting $\pi^{-}$ mesons have been
defined. The obtained results have been compared with the predictions of
CASIMIR.\\
$\bullet$
Using transverse momentum analysis method of P. Danielewicz and G. Odyniec
for the first time the reaction plane has been defined by pions only and the
collective flow effects of $\pi^{-}$ mesons produced in central Mg-Mg
collisions
have been performed. The similar investigation for pions produced in central
C-Ne, C-Cu and O-Pb collisions has been carried out. The dependence of flow
parameter $F$ on $A_{T}$-target mass numbers has been obtained. The obtained results
have been compared with the predictions of QGSM.
\\
$\bullet$
The investigation of the multiparticle azimuthal correlations between
protons and between $\pi^{-}$ mesons in central C-Ne, Mg-Mg, C-Cu and O-Pb
collisions has been carried out. The dependence of
$C(\Delta\varphi)$ - correlation function between particle groups emitted
in the forward and backward hemispheres has been studied for the first time.
By fitting the experimental spectra the asymmetry coefficient $\xi$ and the
strength of correlation $\zeta$ has been determined. The type of correlations 
between protons (C-Ne, C-Cu) and
between $\pi^{-}$ mesons (C-Ne, Mg-Mg, C-Cu, 
O-Pb) and the dependence of the
correlation parameters $\xi$ and $\zeta$ on 
the $A_{T}$-target mass numbers
have been investigated. The obtained results 
have been compared with the
predictions of the QGSM.
\par
\underline{Practical Value}.\\
$\bullet$
The anisotropy coefficient $a$ of $\pi^{-}$ mesons produced in central Mg-Mg
collisions at energy of 3.7 GeV/nucleon increases linearly with
$E_{kin}/E_{kin}^{max}$
starting from approximately 150 MeV. In the range up to 150 MeV the
coefficient is small $a = 0.085 \pm 0.020$, therefore the $\pi^{-}$ mesons are emitted
almost isotropically.
\\
$\bullet$
The shape of the energetic spectra of $\pi^{-}$ mesons emitted at fixed angles in
He(Li,C),  C-Ne,  Mg-Mg, C-Cu, and (C,O)Pb collisions in both laboratory
($10^{o} \leq \theta_{lab.} \leq 180^{o}$) and C.M.
($\theta_{C.M.S.} = 90^{o} \pm 10^{o},  90^{o} \pm 20^{o}$) systems does
not depend on $A_{P}$ and $A_{T}$ for all colliding pairs of nuclei.
The values of the slope parameter $E_{o}$ of the spectra decrease with
increasing of $\theta_{lab}$ from $10^{o}$ to $110^{o}$ and remain
practically constant at $\sim$50-60 MeV for angles larger
than $110^{o}$.
\\
$\bullet$
For $\pi^{-}$  mesons  produced in He(Li,C), C-Ne, Mg-Mg, C-Cu, and
(C,O)Pb collisions the angular
dependence of $E_{o}$ is well described by the fireball model. The temperature
$E_{o}^{\perp}$ and the velocity $\beta$ of the cluster emitting $\pi^{-}$
mesons have been  obtained
from the model. The temperature weakly depends on $A_{P}$, $A_{T}$,
while the velocity does not depend on $A_{P}$, $A_{T}$.
\\
$\bullet$
The obtained collective flow effects and multiparticle azimuthal
correlations in central C-Ne, Mg-Mg, C-Cu and O-Pb collisions at energy
of 3.7 GeV/nucleon, show the persistence of these collective phenomena in
the range of intermediate energy, i.e. from the Bevalac and GSI/SIS up to
Dubna, AGS, RHIC and SPS energies.\\
$\bullet$
The statistical programs of experimental data processing and analyses have
been elaborated for the personal computers in High Energy Physics Institute
of Tbilisi State University, which can be used for the processing and
analysis of the data of the other experiments.
\par
\underline{Structure and Basic Contents of the Thesis}.
The thesis contains an
introduction, five chapter, conclusions and referred bibliography. It
consists of 112 pages, accounts 30 figures and 7 tables. The references
include 147 items.
The experimental data obtained at the Laboratory of High Energies JINR by
the SKM-200--GIBS Collaboration  have been used.
\par
\underline{Introduction} presents the subjects of the thesis, its aims and obtained
results. Structure of the thesis and its short overview are given.
\par
\underline{Chapter I}
is devoted to the description of the spectrometer GIBS, which
has been constructed at the Laboratory of High Energies, at JINR/DUBNA,
and is a modified version of the SKM-200 set-up. The Facility consists of a
2m streamer chamber with the volume of a 2x1x0.6 $m^{3}$, placed in a
magnetic field of $\sim$0.8 T ($\sim$0.9 T for GIBS) and a triggering system. 
In the first
Chapter the main elements of both spectrometers are described and the
physical parameters of spectrometers have been compared. The streamer
chamber was exposed to beams of $^{4}$He, $^{12}$C, $^{16}$O, $^{20}$Ne
(SKM-200) and $^{12}$C, $^{19}$F, $^{24}$Mg (GIBS) nuclei accelerated
in the synchrophasotron up to at energy per incident nucleon of 3.7 GeV.
\par
Solid targets Li, C, Cu and Pb in the form of thin discs with thickness of
(0.2 $\div$ 0.5) g/cm$^{2}$ (the thickness of Li was 1.5 g/cm$^{2}$) were
mounted within the fiducial volume of the SKM-200 chamber. Neon-gas filling
of the chamber also served as a nuclear target. C, F and Mg solid targets
with thickness 0.99 g/cm$^{2}$, 1.34 g/cm$^{2}$ and 1.56 g/cm$^{2}$,
correspondingly, were used for GIBS
set-up. Experimental set-up and the logic of the triggering system are
presented in Fig. 1. The "inelastic" trigger, consisting of two sets of
scintillation counters mounted upstream (S1 -- S4) and downstream (S5, S6)
the chamber, has selected all inelastic interactions of incident nuclei on
a target. The "central" trigger was selecting events defined as those without
charged projectile spectator fragments and spectator neutrons (P/Z $>$ 3 GeV/c),
emitted within a cone of half angle $\theta_{ch} = 2.4^{o}$ or $2.9^{o}$ and
$\theta_{n} = 1.8^{o}$ or $2.8^{o}$
for SKM-200, and $\theta_{ch} = \theta_{n} = 2.4^{o}$ for GIBS set-up.
\par
The streamer chamber pictures were scanned twice at HEPI TSU and JINR/DUBNA.
The final results of scanning and measurement of events have been recorded on 
the Data Summar Tapes
(DST) and then have been analysed by using the standard statistical programs
HBOOK and PAW.
\par
\underline{Chapter II}
is devoted to the description of the intranuclear
cascade model (CAScade Intranuclear Made in Rossendorf) CASIMIR and the Quark
Gluon String Model (QGSM) wich have been used for comparison with the
experimental results presented in this thesis. In CASIMIR model the colliding
nuclei are represented as an assembly of nucleon, the positions and momenta of
which are randomly distributed according to realistic nuclear density and the
law of a cold free Fermi gas, respectively. The nucleon momenta are
distributed according to a sharp Fermi sphere with a radius of $P_{F}$=270 mV/c.
The cascade is subsequently traced in time by the Monte Carlo technique. The
nucleons are assumed to move freely along straight lines between collisions.
For any pair of particles the minimum distance of approach is
\begin{center}
$d_{min} \leq \sqrt{\sigma_{tot}(\sqrt{s})} / \pi , \hspace{2.7cm}$  (1)
\end{center}
where $\sigma_{tot}(\sqrt{s})$ is the total cross section for the pair as a
function of CM energy. The inelastic scattering of nucleons, which is almost
entirely dominated by single- and double-pion production in the energy range
of our interest, is assumed to proceed via the formation of two different
resonances $\Delta(1232)$ and $N^{*}(1640)$.
\par
In the string models the particles are produced via the quark-gluon strings
formations and decays. The quark-gluon strings are excited objects consist of
quark-antiquark or quark-diquark pairs connected via color field. The string
masses are formed as a result of composite quarks longitudinal motion. The
Monte Carlo generator COLLI is based on the Quark Gluon String Model. It is
possible to study hadron-nuclear, hadron-nucleus and nucleus--nucleus
collisions in a wide energy interval and especially in the range of
intermediate energy ($\sqrt{s} \leq$  4 GeV) by this program. The procedure
of event
generation consists of three steps: 1) the definition of configurations of
colliding nucleons, 2) production of quark-gluon strings and 3) fragmentation
of strings (breakup) into observed hadrons. The coordinates of nucleons are
generated according to a realistic nuclear density. It is assumed that the
distance between them is greater than 0.8 fm. The maximum nucleon fermi
momentum is
\begin{center}
$P_{F} =(3\pi^{2})^{1/3} \cdot h \cdot \rho^{1/3}(r) , \hspace{2.7cm}$ (2)
\end{center}
where h=0.197 fm$\cdot$GeV/c, $\rho$ is nuclear density. For main NN and $\pi$N
interactions the following topological quark diagrams were used: binary,
"undeveloped" cylindrical, diffractive and planar. Binary process makes a
main contribution. This reaction predominantly results in the production of
resonances (for instance, $NN \rightarrow N\Delta^{++})$, which are the main 
sources of pions.
The QGSM simplifies the nuclear effects, in particular, coupling of nucleons
inside the nucleus is neglected, and the decay of excited recoil nuclear
fragments and coalescence of nucleons is not included.
\par
\underline{Chapter III}
is devoted to the detailed study of pion production in central
Mg-Mg collisions at energy of E=3.7 GeV/nucl. The choice of $\pi^{-}$
mesons is due to the fact, that they are dominantly produced particles,
carry information about the dynamics of collisions and can be
unambiguously identified. Also, the production of $\pi^{-}$ mesons is
the predominant process at the energies of the
Dubna synchrophasotron. In this chapter the possible sources of experimental
biases and appropriate correction procedures for $\pi^{-}$ mesons
are described. It has been shown that the dependence of the
average kinematical characteristics, $<P_{T}>$ and $<Y>$, of $\pi^{-}$ mesons
on the multiplicity $n_{-} / A_{P}$
differs from that for NN collisions at the same energy, which is due to
nuclear effects. The correlation between $<P_{T}>$ and $Y$ (Fig. 2) has been
studied. One can see that $<P_{T}>$ is smaller in the fragmentation regions of
the projectile and target. The maximum of $<P_{T}>$ corresponds to
$y_{NN}$=1.14. The shape of the $P_{T}$ distributions in various intervals
of $Y$ are described. In the central region of $Y$ the $P_{T}$ spectra are less steep 
and extended up to 1.4 GeV/c, in the fragmentation regions the spectra are more 
sloping and
extended up to (0.5 $\div$ 0.8) GeV/c. The shape of $P_{T}$ dependence
is similar in the fragmentation regions of projectile and target.
\par
The degree of anisotropy in pion emission can be more directly determined by
studying the angular distributions in the c.m.s. of colliding nuclei. The
approximation of this distributions by quadratic form gives for anisotropy
coefficient: $a_{exper}$ = 0.62 $\pm$ 0.02 and $a_{model}$ = 0.63 $\pm$ 0.02
(using the CASIMIR code). It has been shown, that the anisotropy coefficient
increases linearly with
$E_{kin}/E_{kin}^{max}$
($E_{kin}$ is the kinetic energy, $E_{kin}^{max}$ -- the maximum
available energy in the c.m.s. of $\pi^{-}$ mesons) starting from
approximately 150 MeV (Fig. 3). In the range up to 150 MeV the
coefficient is small $a$ = 0.085 $\pm$ 0.020, therefore the $\pi^{-}$ mesons
are emitted almost isotropically, which may be caused by an increase of the
number of NN collisions. The anisotropy coefficient increases linearly with
$E_{kin}/E_{kin}^{max}$
also for the generated data, but the slope is more steep than for the
experimental data.
\par
The energy spectra of $\pi^{-}$ mesons at fixed angles
in the laboratory and c.m. systems in He-Li, He-C, C-Ne, Mg-Mg, C-Cu, C-Pb
and O-Pb collisions at energy of E =3.7 GeV/nucl have been investigated.
The spectra of $\pi^{-}$ mesons depend exponentially on their kinetic energy.
The values of the slope parameter $E_{o}(\theta_{lab.})$ of the spectra - the result
of exponential approximation of energy spectra are presented in Table 1.
One can see that in the forward
hemisphere ($\theta_{lab.} < 90^{o}$) two groups of nucleus pairs can be distinguished:
approximately symmetric and lighter He(Li, C), C-Ne, Mg-Mg and asymmetric
and heavier C-Cu, (C, O)Pb. In each group $E_{o}$ parameters coincide within
errors and, between these two groups, differ no more than 15$\%$. Concerning
$\pi^{-}$ mesons from the backward hemisphere ($\theta_{lab.} > 90^{o}$),
$E_{o}$ parameters coincide within errors for all interactions and there is
no dependence on $A_{P}$, $A_{T}$. The values of the parameter $E_{o}$ in the
c.m. system have the same tendency as in the laboratory system. The shape of
the energy spectra at angular intervals $\theta _{c.m.s.} = 90^{o} \pm 10^{o}$
and $90^{o} \pm 20^{o}$ does not depend on $A_{P}$, $A_{T}$ for all pairs of
nuclei. The values of $E_{o}$ decrease from (250-300 MeV to $\sim$60 MeV with
increasing $\theta_{lab.} > 10^{o}$ from $10^{o}$ to $110^{o}$ (Fig. 4).
From the dependence of $E_{o}$ parameter on $\theta_{lab.} > 10^{o}$
for each colliding pairs of nuclei, the temperature
$E_{o}^{\perp}$ of the cluster formed in the collision and emitting the
particles, and its velosity $\beta$ have been obtained by use of fireball
model approximation
$E_{o} = E_{o}^{\perp} / (1 - \beta cos \theta_{lab.})$.
The values of  $E_{o}^{\perp}$ are comparatively larger $\sim$75 MeV for
approximately symmetric and lighter groups of nuclear pairs
He(Li, C), C-Ne, Mg-Mg, than for asymmetric and heavier pairs of nuclei
C-Cu, (C, O)Pb $\sim$69 MeV.
\par
\underline{Chapter IV}
is devoted to the experimental results obtained from the in-plane
transverse momentum analysis of $\pi^{-}$ mesons in central C-Ne, Mg-Mg,
C-Cu and O-Pb interactions at energy of E=3.7 GeV/nucl. As the $\pi^{-}$
mesons are emitted mainly from decays of $\Delta$-isobars (at least
$\sim 80 \%$), we decided to investigate
whether a single $\pi^{-}$ meson of the event knows something about
its origin and the question whether they are collectively correlated.
Many different methods were proposed for the study of flow effects
in relativistic nuclear collisions, of which the most commonly used
are - the transverse momentum analysis technique developed by
P. Danielewicz and G. Odyniec and Fourier
analysis of the azimuthal distributions on the event-by-event basis in
relatively narrow rapidity windows proposed by M. Demoulins, S. Voloshin and
Y. Zhang. We have used the technique of P.Danielewicz and G. Odyniec in our
analysis. The method involves two basic ideas: 1) to select the rapidity
range and rapidity dependent waiting factors to the real reaction plane, and
2) to remove trivial and spurious self-correlations from the projections.
Autocorrelations are removed by calculating $\overrightarrow{Q}$
individually for each particle without including that particle into sum:
\begin{center}
$\overrightarrow{Q_{j}}$=$\sum\limits_{i\not=j}\limits^{n}$$\omega_{i}$
$\overrightarrow{P_{{T}i}}$ \hspace{6.7cm}  (3)
\end{center}
where i is a particle index and $\omega_{i}$ is a weight, taken as 1 for
y$_{i}$$>$ y$_{cms}$ and -1 for y$_{i}$$<$ y$_{cms}$
(y$_{cms}$ is c.m.s. rapidity and y$_{i}$ is the rapidity of
particle $i$). The transverse momentum of each particle in the estimated
reaction plane is calculated as:
\begin{center}
$P_{xj}\hspace{0.01cm}^{\prime} =
\sum\limits_{i\not=j}\limits^{n} \omega_{i} \cdot (
\overrightarrow{P_{{T}j}} \cdot \overrightarrow{P_{{T}i}})
/ \vert\overrightarrow{Q}\vert$ \hspace{4.1cm}  (4)
\end{center}
The $\overrightarrow{Q}$ vector have been constructed from the
$\overrightarrow{P_{{T}i}}$ of only $\pi^{-}$ mesons event-by-
event for the events with multiplicity $n_{-} >$ 7 in Mg-Mg collisions.
It is known, that the estimated reaction plane differs from the
true one, due to the finite number of particles in each event.
The component $P_{x}$ in the true reaction plane is systematically
larger then $P_{x}\hspace{0.01cm}^{\prime}$ the component in the
estimated plane, hence
$<P_{x}>=<P_{x}\hspace{0.01cm}^{\prime}>/<cos \varphi>$,
where $\varphi$ is the angle between the
estimated and true planes and the correction factor $K=1/<cos \varphi>$.
According to the method of Danielewicz and Odyniec, for the definition of
$<cos \varphi>$, we divided each event randomly into two equal
sub-events, constructed vectors
$\overrightarrow{Q_{I}}$ and $\overrightarrow{Q_{II}}$
and estimated azimuthal angle $\phi_{1, 2}$ between these two vectors
$<cos \varphi>=<cos \varphi_{1, 2}>$. The correction factor, averaged over all the
multiplicities, $K$ = 1.51 $\pm$ 0.05 has been obtained. Figure 5 shows the
dependence of the estimated
$<P_{x}\hspace{0.01cm}^{\prime}(Y)>$ on $Y$ for
pions produced in the central Mg-Mg collisions. The data exhibit S-shape
behavior similar to the form of
the $<P_{x}>$ spectra for protons and pions obtained at lower energies and
identified as nucleon and pion collective flow. The slope at midrapidity has
been extracted from a linear fit of the data for $Y$ between 0.2 $\div$ 2,
the value of $F$ = 48 $\pm$ 5 MeV/c have been obtained. In the model calculations
the reaction plane is known a priori and is refered as the reaction plane.
For simulated events (QGSM) the component in the true reaction plane
$P_{x}$ has been calculated and the dependence of
$<P_{x}\hspace{0.01cm}^{\prime}(Y)>$ on $Y$
has been obtained for both not fixed $\tilde{b}$ (6225 events) and fixed
$<b>$=1.34 fm (6212 events) impact parameters
(Fig. 5). The experimental and QGSM results coincide within errors. The
values of  $F$, obtained from the QGSM are: $F$ = 53 $\pm$ 3 (MeV/c) - for not
fixed $\tilde{b}$, and $F$ = 51 $\pm$ 4 MeV/c - for 1.34 fm.
\par
The similar in-plane analysis has been performed on $\pi^{-}$ mesons too
in central C-Ne, C-Cu and O-Pb collisions at the same energy. In this case
the reaction plane has been obtained by only $\pi^{-}$ mesons. The observed
dependences show the typical S-shape behavior reflecting the presence of flow effects. 
The values
of flow $F$ increases with the mass number of target $A_{T}$ from
33 $\pm$ 4 (MeV/c) to 50 $\pm$ 6 (MeV/c)  (see Table 2). The obtained
experimental results in central  C-Ne, C-Cu and O-Pb collisions have been
compared with the predictions of the QGSM. For this purpose C-Ne (6272),
C-Cu (9327) and O-Pb (2431) collisions have been generated for impact
parameters b=2.20, 2.75 and 3.74 (fm), respectively. The Quark Gluon String
Model reproduces the $<P_{x}>$ distributions and satisfactorily describes the
obtained experimental results  for all pairs of nuclei.
\par
It seems, that the obtained results indicate that the flow behavior of
$\pi^{-}$ mesons is due both to the flow of $\Delta$ resonances and of the
nuclear shadowing effect. Our experimental results, obtained using the
streamer chamber technique, provide quantitative information on the
directed flow of $\pi^{-}$ mesons and their dependence on target-mass, 
complementing the
experimental data available from the BEVALAC, GSI-SIS, AGS and CERN/SPS.
\par
\underline{Chapter V}
is devoted to the multiparticle azimuthal correlations between
protons and between
pions in central C-Ne, Mg-Mg, C-Cu and O-Pb collisions at energy of
3.7 GeV/nucleon. Identification of secondary particles becomes very
important for the study of products of nucleus-nucleus ($A_{P}$ - $A_{T}$) collisions
via collective variables, which are defined event by event. The admixture of
$\pi^{-}$ mesons amongst  the charged positive particles is about
(25 $\div$ 27). The statistical method has been used for the identification
of $\pi^{+}$ mesons. The main assumption is based on the similarity of
spectra of $\pi^{-}$ and  $\pi^{+}$ mesons
($n_{-}, P_{T},  P_{L}$). The two-dimensional distribution of
$P_{T}$, $P_{L}$ variables have been used for identification of $\pi^{-}$
mesons. After performed identification
the admixture of $\pi^{-}$ mesons amongst the protons is not exceeding
(5 $\div$ 7)$\%$. The azimuthal correlation function C($\Delta \varphi$)
was defined by the relative angle between the transverse momentum vectors
sums of particles emitted in forward and backward hemispheres:
\begin{center}
$\overrightarrow{Q_{F}} = \sum\limits_{y_{i}\geq<y>} \overrightarrow{P_
{T i}} \hspace{6.3cm} $   (5)
\end{center}
and
\begin{center}
$\overrightarrow{Q_{B}}=\sum\limits_{y_{i}<<y>}\overrightarrow{P_
{T i}} \hspace{6.5cm}$    (6)
\end{center}
where $<y>$ is the average rapidity in each event.
The correlation function C($\Delta \varphi$) is constructed as follows:
\begin{center}
$C(\Delta \varphi$) = dN/d$\Delta \varphi$  \hspace{6cm}    (7)
\end{center}
where $\Delta \varphi$ is the angle between these vectors:
\begin{center}
$\Delta \varphi = arccos(\overrightarrow{Q_{B}}\cdot
\overrightarrow{Q_{F}}) /
\vert \overrightarrow{Q_{B}} \vert \cdot \vert \overrightarrow{Q_{F}} \vert.
\hspace{3cm} $    (8)
\end{center}
For protons a "back-to-back" negative correlations were observed in C-Ne and
C-Cu collisions i.e., protons are preferentially emitted at
$\Delta \varphi$ = 180 $^{o}$. To quantify these experimental results, the
data were fitted by
$C(\Delta \varphi$) = 1 +$\xi cos(\Delta \varphi$), where $\xi$ is the
asymmetry coefficient. The strength of correlation $\zeta$ was defined as
\begin{center}
$\zeta = C(0^{o})/C(180^{o}) = (1 + \xi)/(1 -
\xi).  \hspace{3.1cm} $  (9)
\end{center}
In Table 3 the asymmetry coefficient $\xi <$ 0
and the strength of correlation $\zeta <$ 1 for protons in both
C-Ne and C-Cu interactions are presented. One can
see that $\vert \xi \vert$ increases and $\zeta$ decreases
with increase of target mass. A back-to-back  pion  correlations  have
been  obtained  for a light, symmetric  pairs  of  nuclei  C-Ne  and  Mg-Mg
($n_{-} \geq 7$),  where $\xi$ and $\zeta$ parameters have the same
behavior as for protons (Fig. 6a). For heavy pairs of  nuclei C-Ne and O-Pb,
side-by-side (positive) pion correlations were observed i.e. in this case
they are preferentially emitted at $\Delta \varphi = 0^{o}$ (Fig. 6b).
The asymmetry coefficient $\xi$ and the strength of
correlations $\zeta$ increase with increase of $A_{P}$ and $A_{T}$. The
variation of pion correlations (from negative for light
systems of C-Ne, Mg-Mg to positive for heavy systems of nuclei C-Cu, O-Pb),
which have been observed in our experiment, indicate, that at Dubna energies
the flow behavior of $\pi^{-}$ mesons in a light system is due the flow of
$\Delta$ resonances ($\pi N \rightarrow \Delta$ and
$\Delta N \rightarrow NN $), whereas the antiflow behavior in a heavier
system is the result of the nuclear shadowing effect.
\par
To be convinced that the azimuthal correlations between protons and between
pions are due to correlations between these particles and can not be the
result of detector biases or finite-multiplicity effects, we obtained the
of $C(\Delta \psi$) on $\psi$
for secondaries, where $\psi$ is the angle between the
transverse momentum of each particle emitted in the backward (forward)
hemisphere and $\overrightarrow{Q_{F}}(\overrightarrow{Q_{B}})$ vector,
respectively. No correlations have been
obtained for C-Cu interactions both for protons and for pions. Similar
results have been obtained for C-Ne collisions.
\par
In order to extend these investigations, the relation between
$<P_{x}\hspace{0.01cm}^{2}>$ and the angle
$\Delta \varphi$, where $\Delta \varphi$
is the opening angle between
$\overrightarrow{Q_{F}}$ and $\overrightarrow{Q_{B}}$
vectors has been obtained. One can
see from Fig. 7, that for protons in C-Ne and C-Cu collisions the
distributions show S-shape behavior and slopes of distributions increase
with target mass; We stress, that at $\Delta \varphi = 90^{o}$ the values of
$<P_{x}>$ do not depend on $A_{T}$.
\par
The Quark Gluon String Model (QGSM) satisfactorily describes the obtained
results for protons and  for $\pi_{-}$ mesons for all pairs of nuclei.
\par
The comparison of our experimental results with the data of various
collaborations (BEVALAC, JINR, CERN/SPS) has been carried out and had been
showed that these multiparticle azimuthal correlations between the same
types of secondaries exist for the energy interval of
(0.1 $\div$ 200) GeV/nucleon.
\par
\underline{Conclusions} list main results of the conducted research work,
which are outlined bellows.
\par
\underline {The Main Results of the Thesis}
\\
1. A detailed study of pion production in central Mg-Mg collisions at energy
of E=3.7 GeV per nucleon (P = 4.3 GeV/c/nucl) have been carried by the
Lorentz invariant variables: the  rapidity Y and the transverse
momentum $P_{T}$. The average values of $<Y>$ and $<P_{T}>$ decrease
with increasing of multiplicity,
$n_{-}/A_{P}$, different to NN collisions, where the values of both average
kinematical characteristics remain constant with increase of $n_{-}/A_{P}$.
$n_{-}/A_{P}$ is the measure of the impact parameter $b$.
\par
The correlation between $<P_{T}>$ and $Y$ has been studied: $<P_{T}>$
is smaller in the fragmentation regions of the projectile and target
(the maximum of $<P_{T}>$ corresponds to $y_{NN}$ = 1.14).
In the central region of $Y$ the $P_{T}$ distributions are
less steep and extended up to 1.4 GeV/c.
\\
2. In the analysis of angular distributions of $\pi^{-}$ mesons in Mg-Mg
collision the anisotropy coefficient $a_{exper.}$ = 0.62 $\pm$ 0.02 has been obtained;
The obtained results qualitatively agree with the predictions of the
intranuclear cascade model CASIMIR $a_{model}$ = 0.63 $\pm$ 0.02.
The anisotropy coefficient increases linearly with
$E_{kin}/E_{kin}^{max}$
($E_{kin}$ is the kinetic energy,
$E_{kin}^{max}$ -- the maximum available 
energy in the c.m.s. of
$\pi^{-}$ mesons) starting from approximately 150 MeV. In the range up to 150 MeV
the coefficient  is small $a$ = 0.085 $\pm$ 0.020, therefore the $\pi^{-}$
mesons are emitted almost isotropically. The anisotropy coefficient
increases linearly with
$E_{kin}/E_{kin}^{max}$
also for the generated data (CASIMIR), but the slope is
more steep than for the experimental data.
\\
3. The energy spectra of $\pi^{-}$ mesons emitted at fixed angles in the
laboratory ($10^{o} \leq \theta_{lab} \leq 90^{o}$) and c.m. systems
($\theta_{c.m.s.} = 90^{o} \pm 10^{o}$ and $90^{o} \pm 20^{o}$) in He-Li,
He-C, C-Ne, Mg-Mg, C-Cu, C-Pb and O-Pb collisions collisions at energy of
E=3.7 GeV/nucl depend exponentially on their kinetic energy. The shape of
spectra in both systems does not depend on $A_{P}$, $A_{T}$ for all
colliding pairs of nuclei. The values of the slope parameter $E_{o}$ of
the spectra depend on $A_{P}$, $A_{T}$ for angles $\theta_{lab} < 90^{o}$
and are independent for $\theta_{lab} > 90^{o}$. The values of $E_{o}$
decrease from $\sim$250-300 MeV to $\sim$60 MeV with increasing
$\theta_{lab}$ from $10^{o}$ to $110^{o}$,
and remain practically constant at $\sim$50-60 MeV for angles larger
$110^{o}$. The angular dependence of $E_{o}$ parameter has been approximated
by the fireball model. From the approximation the parameters of the model
- the temperature $E_{o}^{\perp}$ and velosity $\beta$ of the cluster
emitting $\pi^{-}$ mesons have been extracted.
The temperature weakly depends on $A_{P}$, $A_{T}$ while the velocity does not depend
on $A_{P}$, $A_{T}$.\\
4. The investigation of in-plane flow effects for $\pi^{-}$ mesons
in central Mg-Mg collisions by the transverse momentum technique of
P. Danielewicz and G. Odyniec has been carried out. For the first time,
the reaction plane heve been determined only by $\pi^{-}$ mesons in each event. 
The similar analysis of in-plane flow effects for $\pi^{-}$ mesons production 
in central C-Ne, C-Cu
and O-Pb collisions at the same energy has been performed. The observed dependence of
$<P_{x}\hspace{0.01cm}^{\prime}(Y)>$ on $Y$
shows S-shape behavior. The values of $F$ flow parameter,
defined
as the slope at midrapidity, have been extracted for all colliding pairs of
nuclei. $F$ increases with the mass numbers of target $A_{T}$ from
33 $\pm$ 4 (MeV/c) to 50 $\pm$ 6 (MeV/c). The Quark Gluon String Model
(QGSM) reproduces the $<P_{x}>$ distributions and satisfactorily describes
obtained experimental results for all pairs of nuclei.
\\
5. The multiparticle azimuthal correlations between protons and between
pions in central C-Ne, Mg-Mg, C-Cu and O-Pb collisions have been studied.
The dependence of the azimuthal correlation function $C(\Delta \varphi$)
on the $\Delta \varphi$, where $\Delta \varphi$, is angle between the
transverse momentum vectors sums of particles emitted in forward and
backward hemispheres, in these collisions have been investigated:
\par
For protons a "back-to- back" (negative) correlations were observed
in central C-Ne and C-Cu interactions. The asymmetry coefficient
$\vert \xi$ ($\xi < 0$) increases and the strength of correlation
$\zeta$ ($\zeta < 1$) decreases with increase of the target mass $A_{T}$.
\par
"Back-to-back" pion correlations have been obtained for a light pairs
of nuclei (C-Ne and Mg-Mg), where $\xi$ and $\zeta$ parameters have the
same behavior as for protons. For heavy pairs of nuclei
(O-Pb and C-Cu) "side-by-side" (positive) pions correlations were observed.
The asymmetry  ($\xi > 0$) and strength of correlation ($\zeta > 1$)
increase with increase of $A_{P}$ and $A_{T}$.
\par
The QGSM satisfactorily describes multiparticle azimuthal correlations
of protons and $\pi^{-}$ mesons for all pairs of nuclei.
\par
\underline{Approbation of the Work}. The basic results of the
investigations were reported at: Seminars in High Energy Physics Institute
of Tbilisi State University and in Dubna Joint Institute of Nuclear Research,
International Symposium in Elemantary Particle Physics dediceated to the
memory G. Chikovani, 21-23 September, 1998, Tbilisi; On the 1999 European
School of young scientist in High-Energy Physics, Casta-Papiernicka,
Slovac Republic, 22 August - 4 September and on the Seventh International
Conference in Nucleus-Nucleus Collisions 2000, July 3-7, Strasbourg, France.
\  \par
\  \par
\  \par
\  \par
\  \par
\underline{Publications}
\  \par
\  \par
1. Chkhaidze L., Djobava T. and Kharkhelauri L., "The investigation 
of energy spectra of $\pi^{-}$ mesons at fixed angles in nucleus-nucleus 
interactions at a momentum of 4.5 GeV/c per incident nucleon." 
-- J. Phys., 1993,  v.19G, p.1155-1161.
\par
2. Chkhaidze L., Djobava T., Gogiberidze G., Kharkhelauri L. and 
Mosidze M., "Study of the inclusive reaction  Mg-Mg $\rightarrow$ $\pi^{-}$+X 
at a momentum of 4.3 GeV/c per incident nucleon."  
-- J. Phys., 1996, v.22G,  p.641-651.
\par
3. Chkhaidze L., Djobava T., Kharkhelauri L. and Mosidze M. "The 
comparison of characteristics  of $\pi^{-}$ mesons produced in central Mg-Mg 
interactions with the quark gluon string model predictions."
-- Eur. Phys. J., 1998, v.1A, p.299-306;
hep-ex/980528;
\par
4. Chkhaidze L., Djobava T and Kharkhelauri L., "Multiparticle azimuthal 
correlations in central nucleus-nucleus collisions at energy of 3.7 GeV per
 nucleon." --  Bulletin of the Georgian Academy of Sciences "Moambe", 2001, 
v.164, N1, p.51-55.
\par
5. Chkhaidze L., Djobava T. and Kharkhelauri L., "Study of multiparticle azimuthal 
correlations in central CNe, MgMg, CCu and OPb interactions at energy of 3.7 GeV 
per nucleon."
-- Phys. Atom. Nucl. 65:1479-1486, 2002; Yad. Fiz. 65: 1515-1522;
nucl-ex/0107017.
\newpage
Table 1. The values of the paramether E$_{o}$ -- the result of approximating
$\pi_{-}$ mesons energy spectra by formula 
$\sigma_{inv.}$ = Aexp(E$_{kin}$/E$_{o}$).\\
\   \par
\   \par
\   \par
\   \par
$\hspace{15.9cm}$
\begin{tabular}{|c|c|c|c|c|c|c|}     \hline
\multicolumn{1}{|c|}{}&
\multicolumn{1}{|c|}{}&
\multicolumn{5}{|c|}{}\\
\multicolumn{1}{|c|}{$\theta_{lab}$}&
\multicolumn{1}{|c|}{$\Delta$ $E_{kin}$}&
\multicolumn{5}{|c|}{$E_{o}$ (GeV)}\\
\multicolumn{1}{|c|}{intervals}&
\multicolumn{1}{|c|}{(GeV)}&
\multicolumn{5}{|c|}{}\\
\cline{3-7}
\multicolumn{1}{|c|}{(grad)}&
\multicolumn{1}{|c|}{}&
\multicolumn{1}{|c|}{He(Li,C)}&
\multicolumn{1}{|c|}{C-Ne}&
\multicolumn{1}{|c|}{Mg-Mg}&
\multicolumn{1}{|c|}{C-Cu}&
\multicolumn{1}{|c|}{(C,O)Pb}\\
\hline
\multicolumn{1}{|c|}{10-20}&
\multicolumn{1}{|c|}{0.2-0.3}&
\multicolumn{1}{|c|}{0.408 $\pm$ 0.009}&
\multicolumn{1}{|c|}{0.390 $\pm$ 0.010}&
\multicolumn{1}{|c|}{0.378 $\pm$ 0.006}&
\multicolumn{1}{|c|}{0.349 $\pm$ 0.008}&
\multicolumn{1}{|c|}{0.318 $\pm$ 0.008}\\
\hline
\multicolumn{1}{|c|}{20-30}&
\multicolumn{1}{|c|}{0.15-2.1}&
\multicolumn{1}{|c|}{0.289 $\pm$ 0.007}&
\multicolumn{1}{|c|}{0.283 $\pm$ 0.009}&
\multicolumn{1}{|c|}{0.271 $\pm$ 0.005}&
\multicolumn{1}{|c|}{0.259 $\pm$ 0.006}&
\multicolumn{1}{|c|}{0.234 $\pm$ 0.007}\\
\hline
\multicolumn{1}{|c|}{30-40}&
\multicolumn{1}{|c|}{0.10-1.6}&
\multicolumn{1}{|c|}{0.211 $\pm$ 0.006}&
\multicolumn{1}{|c|}{0.200 $\pm$ 0.007}&
\multicolumn{1}{|c|}{0.210 $\pm$ 0.005}&
\multicolumn{1}{|c|}{0.182 $\pm$ 0.004}&
\multicolumn{1}{|c|}{0.175 $\pm$ 0.005}\\
\hline
\multicolumn{1}{|c|}{40-50}&
\multicolumn{1}{|c|}{0.05-1.2}&
\multicolumn{1}{|c|}{0.161 $\pm$ 0.005}&
\multicolumn{1}{|c|}{0.159 $\pm$ 0.006}&
\multicolumn{1}{|c|}{0.170 $\pm$ 0.004}&
\multicolumn{1}{|c|}{0.140 $\pm$ 0.004}&
\multicolumn{1}{|c|}{0.136 $\pm$ 0.004}\\
\hline
\multicolumn{1}{|c|}{50-70}&
\multicolumn{1}{|c|}{0.05-1.0}&
\multicolumn{1}{|c|}{0.123 $\pm$ 0.004}&
\multicolumn{1}{|c|}{0.121 $\pm$ 0.006}&
\multicolumn{1}{|c|}{0.122 $\pm$ 0.003}&
\multicolumn{1}{|c|}{0.109 $\pm$ 0.003}&
\multicolumn{1}{|c|}{0.098 $\pm$ 0.003}\\
\hline
\multicolumn{1}{|c|}{70-90}&
\multicolumn{1}{|c|}{0.05-0.7}&
\multicolumn{1}{|c|}{0.083 $\pm$ 0.005}&
\multicolumn{1}{|c|}{0.080 $\pm$ 0.005}&
\multicolumn{1}{|c|}{0.089 $\pm$ 0.004}&
\multicolumn{1}{|c|}{0.079 $\pm$ 0.003}&
\multicolumn{1}{|c|}{0.071 $\pm$ 0.003}\\
\hline
\multicolumn{1}{|c|}{90-110}&
\multicolumn{1}{|c|}{0.05-0.7}&
\multicolumn{1}{|c|}{0.059 $\pm$ 0.005}&
\multicolumn{1}{|c|}{0.057 $\pm$ 0.005}&
\multicolumn{1}{|c|}{0.059 $\pm$ 0.002}&
\multicolumn{1}{|c|}{0.059 $\pm$ 0.002}&
\multicolumn{1}{|c|}{0.058 $\pm$ 0.002}\\
\hline
\multicolumn{1}{|c|}{110-130}&
\multicolumn{1}{|c|}{0.05-0.5}&
\multicolumn{1}{|c|}{0.064 $\pm$ 0.011}&
\multicolumn{1}{|c|}{0.060 $\pm$ 0.008}&
\multicolumn{1}{|c|}{0.053 $\pm$ 0.005}&
\multicolumn{1}{|c|}{0.052 $\pm$ 0.003}&
\multicolumn{1}{|c|}{0.055 $\pm$ 0.003}\\
\hline
\multicolumn{1}{|c|}{130-180}&
\multicolumn{1}{|c|}{0.05-0.5}&
\multicolumn{1}{|c|}{0.046 $\pm$ 0.004}&
\multicolumn{1}{|c|}{0.052 $\pm$ 0.007}&
\multicolumn{1}{|c|}{0.052 $\pm$ 0.004}&
\multicolumn{1}{|c|}{0.046 $\pm$ 0.003}&
\multicolumn{1}{|c|}{0.041 $\pm$ 0.003}\\
\hline
\end{tabular}
\newpage
Table 2. The number of experimental events $N_{event}$, the
average multiplicity of protons (participant) $<n_{prot}>$
and $\pi^{-}$ mesons $<n_{\pi_{-}}>$, and the flow parameter in
central C-Ne, Mg-Mg, C-Cu and O-Pb collisions (the reaction
plane has been obtained by $\pi^{-}$ mesons only).
\   \par
\   \par
\   \par
\   \par
\   \par
\   \par
\begin{tabular}{|l|c|c|c|c|}    \hline
&&&&   \\
$A_{P}$ - $A_{T}$ & C-Ne & Mg-Mg & C-Cu & O-Pb \\
&&&&   \\
\hline
$N_{event}$ & 723 & 6239 & 1866 & 732 \\
\hline
$<n_{prot}>$ & 12.4 $\pm$ 0.5 & -- & 19.5 $\pm$ 0.6 & -- \\
\hline
$<n_{\pi_{-}}>$ & 7.8 $\pm$ 0.4 & 8.2 $\pm$ 0.4 & 6.6 $\pm$ 0.3 & 9.8 $\pm$ 0.5  \\
\hline
$F$ (MeV/c) & 33 $\pm$ 4 & 48 $\pm$ 5 & 41 $\pm$ 5 & 50 $\pm$ 6  \\
\hline
\end{tabular}
\newpage
Table 3. The number of experimental events $N_{\rm{event}}$ and
of participant protons $N_{\rm{prot.}}$, the asymmetry coefficient
($\rm{\xi}$), the strength of the correlation ($\rm{\zeta}$) and the average
rapidity ($\rm{<y>}$) of protons in central C-Ne and C-Cu collisions.\\
\   \par
\   \par
\   \par
\   \par
\   \par
\   \par
\begin{tabular}{|c|c|c|c|c|c|}    \hline
&    &     &    &     &   \\
$A_{P}$ - $A_{T}$ & $N_{event}$ & $N_{prot.}$ &
$\xi$ & $\zeta$ & $<y>$ \\
\hline
&    &     &    &     &   \\
C-Ne  & 723 &  9201  & -0.23 $\pm$ 0.05 & 0.63 $\pm$ 0.09 & 1.07
$\pm$ 0.07\\
\hline
&    &     &    &     &   \\
C-Cu  & 663 & 12715  & -0.35 $\pm$ 0.05 & 0.48 $\pm$ 0.06 & 0.73
$\pm$ 0.05\\
\hline
\end{tabular}
\newpage
\begin{figure} \begin{center}
\epsfig{file=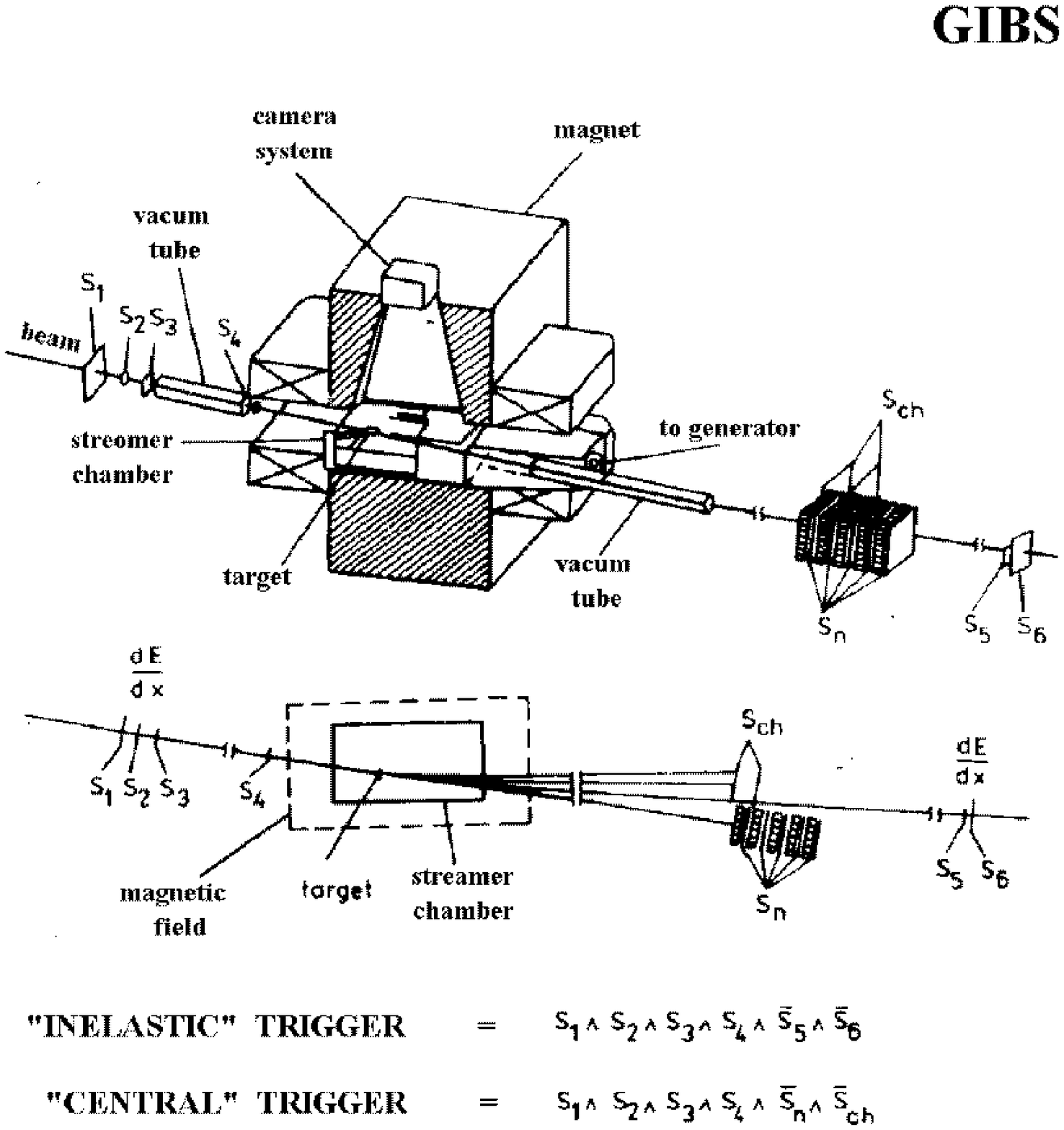,bbllx=0pt,bblly=0pt,bburx=594pt,bbury=842pt,
width=18.0cm,angle=0}
\end{center}
\vspace{-4.9cm}
\hspace{0.cm}
\begin{minipage}{16.0cm}
\caption
{ Experimental set-up. The trigger and trigger distances are
not to scale.}
\end{minipage}
\end{figure}
\pagebreak
\begin{figure}
\begin{center}
\epsfig{file=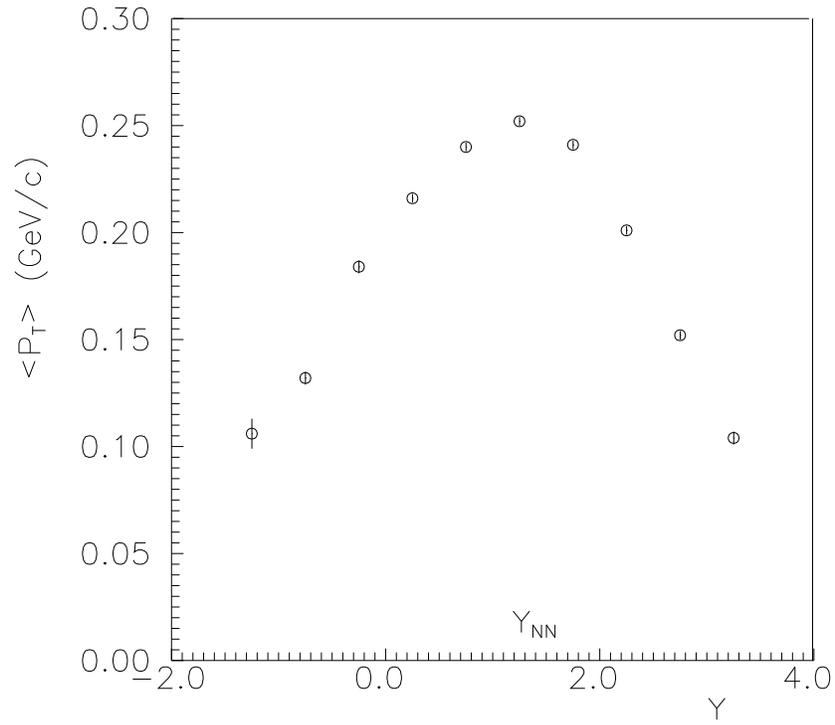,bbllx=0pt,bblly=0pt,bburx=594pt,bbury=842pt,
width=18cm,angle=0}
\end{center}
\vspace{-10.cm}
\hspace{3.cm}
\begin{minipage}{13.0cm}
\caption
{The dependence of the average transverse momentum $< P_{T} >$ on the
rapidity $Y$ for $\pi^{-}$ mesoms.}
\end{minipage}
\end{figure}
\pagebreak
\begin{figure}
\begin{center}
\epsfig{file=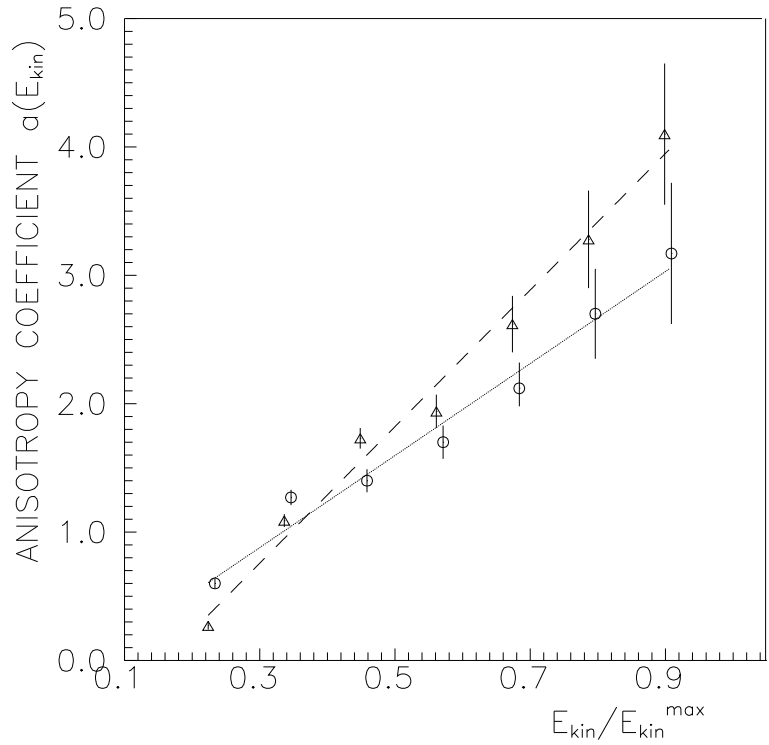,bbllx=0pt,bblly=0pt,bburx=594pt,bbury=842pt,
width=18cm,angle=0}
\end{center}
\vspace{-10.cm}
\hspace{3.cm}
\begin{minipage}{13.0cm}
\caption
{The dependence of the anisotropy coefficient $a$ on the 
$E_{kin} / E_{kin}^{max}$ 
($E_{kin}$ is the kinetic energy, $E_{kin}^{max}$
-- maximum available energy in the c.m.s. of $\pi^{-}$ mesons):
$\circ$ -- experiment, $\bigtriangleup$ --  CASIMIR. The lines
are the result of a linear approximation of the experimental
and generated data, respectively.}
\end{minipage}
\end{figure}
\pagebreak
\begin{figure}
\begin{center}
\epsfig{file=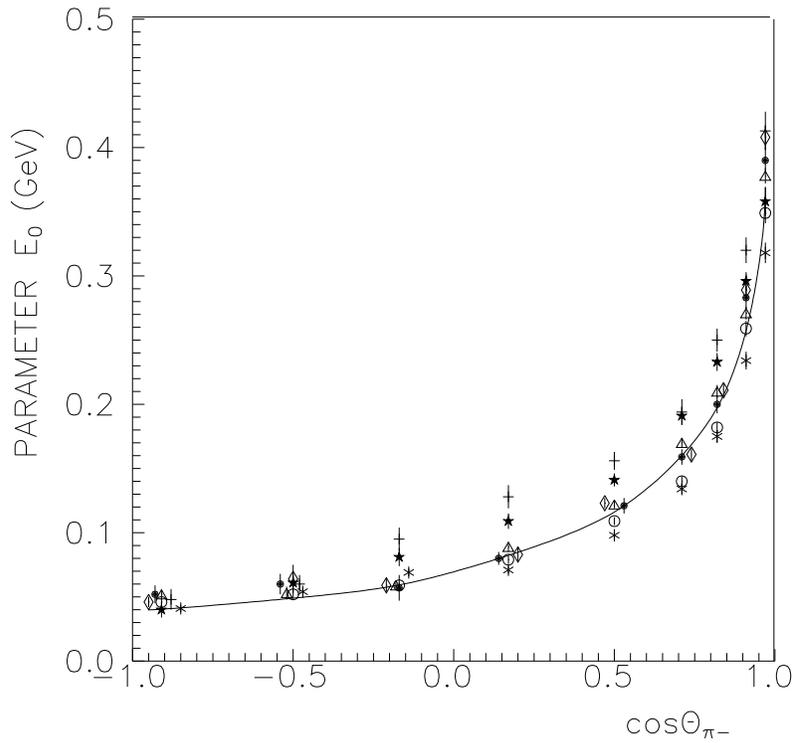,bbllx=0pt,bblly=0pt,bburx=594pt,bbury=842pt,
width=18cm,angle=0}
\end{center}
\vspace{-10.cm}
\hspace{3.cm}
\begin{minipage}{13.0cm}
\caption
{The angular dependence of the parameter $E_{o}$ for
$\bigtriangleup$ -- He(Li, C),
$\ast$ -- C-Ne,
$\bullet$  -- Mg-Mg, 
$\circ$ -- C-Cu, $\diamond$ -- (C,O)Pb
collisions (P=4.5 GeV/c/nucl) and $\star$ -- He-C, 
+ -- C-C collisions (P=4.2 GeV/c/nucl). The solid curve is the
result of approximating the averaged values $E(\theta_{lab})$ by
function $E_{o}=E_{o}^{\perp}$ / (1- $\beta$ cos $\theta_{lab}$)}.
\end{minipage}
\end{figure}
\pagebreak
\begin{figure}
\begin{center}
\epsfig{file=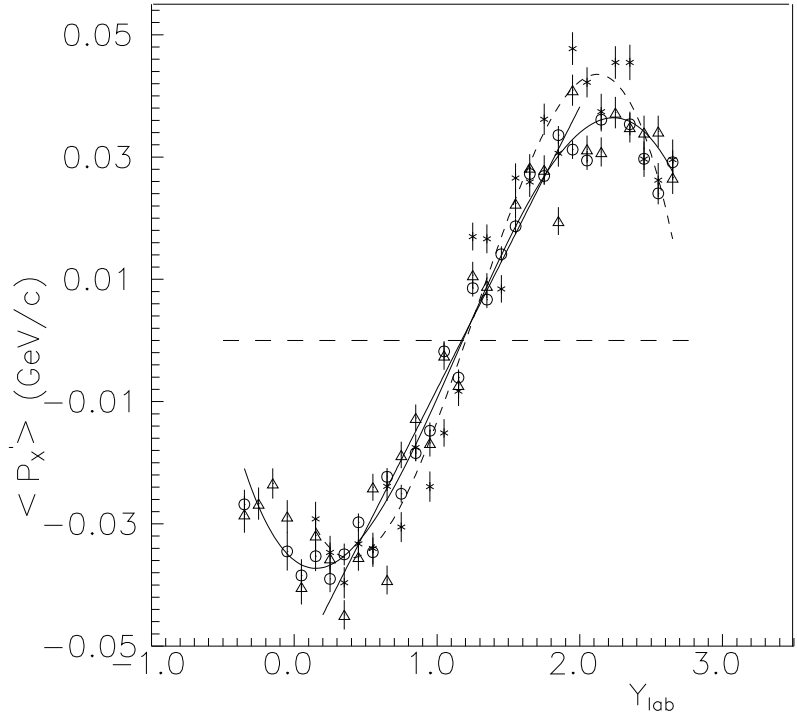,bbllx=0pt,bblly=0pt,bburx=594pt,bbury=842pt,
width=18cm,angle=0}
\end{center}
\vspace{-10.cm}
\hspace{3.cm}
\begin{minipage}{13.0cm}
\caption
{The dependence of $< P_{x}\hspace{0.01cm}^{\prime}(Y) >$ on $Y_{lab}$.
for $\pi^{-}$ mesons produced in the central Mg-Mg collisions:
$\circ$ -- the experimental data and QGSM generated data for both
$\bigtriangleup$ -- fixed $b$= 1.34 fm and
$\ast$ -- not fixed $\tilde{b}$ impact parameters.
The solid line is the result of the the linear approximation of 
experimental data in the interval of
$Y$ - 0.2 $\div$ 2.0. The curves for
visual presentation of the QGSM events (solid - for fixed $b$,
dashed -for $\tilde{b}$) 
-- result of approximation by 4-th order polynomial.}
\end{minipage}
\end{figure}
\pagebreak
\begin{figure}
\begin{center}
\epsfig{file=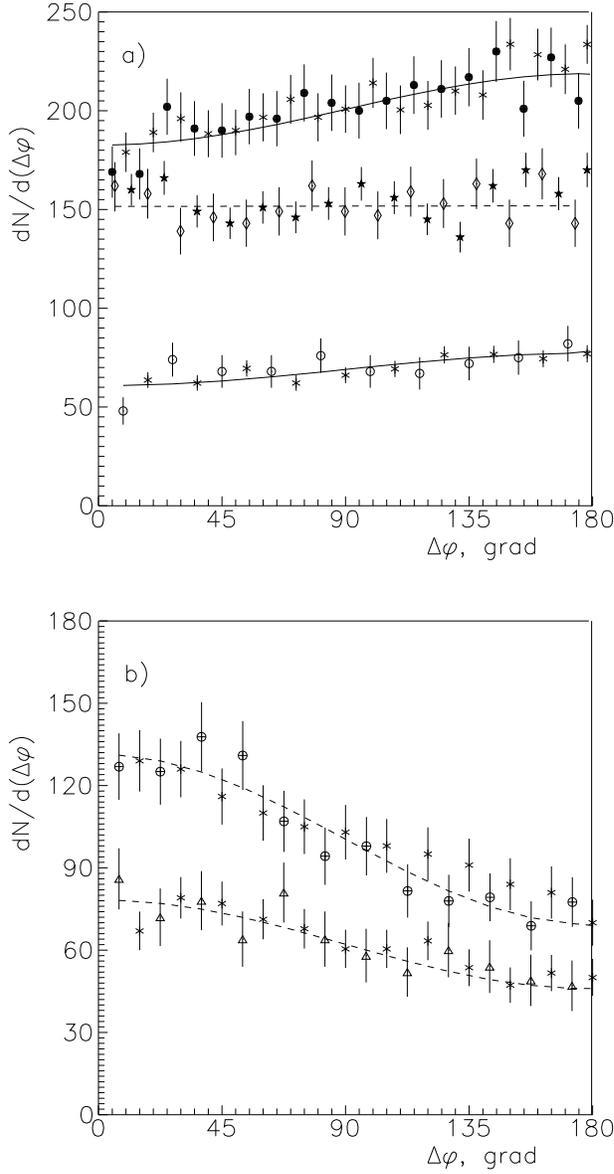,bbllx=0pt,bblly=0pt,bburx=594pt,bbury=842pt,
width=18cm,angle=0}
\end{center}
\vspace{-4.9cm}
\hspace{0.cm}
\begin{minipage}{16.0cm}
\caption
{The correlation function $C(\Delta \varphi$) as descreibed in the
text for $\pi^{-}$-mesons from:
(a) $\circ$ -- C-Ne, $\bullet$ -- Mg-Mg ($n_{-} \geq$ 7),
 $\diamond$ -- Mg-Mg ($n_{-} <$ 7) and (b) $\bigtriangleup$ -- C-Cu,
$\oplus$ -- O-Pb collisions.
{\scriptsize$\times$}, $\star$ -- the QGSM generated data,
respectively. The solid and dashed curves -- results of the approximation
of the data by function  $C(\Delta\varphi$)=1 + $\xi$cos($\Delta\varphi$).}
\end{minipage}
\end{figure}
\pagebreak
\begin{figure}
\begin{center}
\epsfig{file=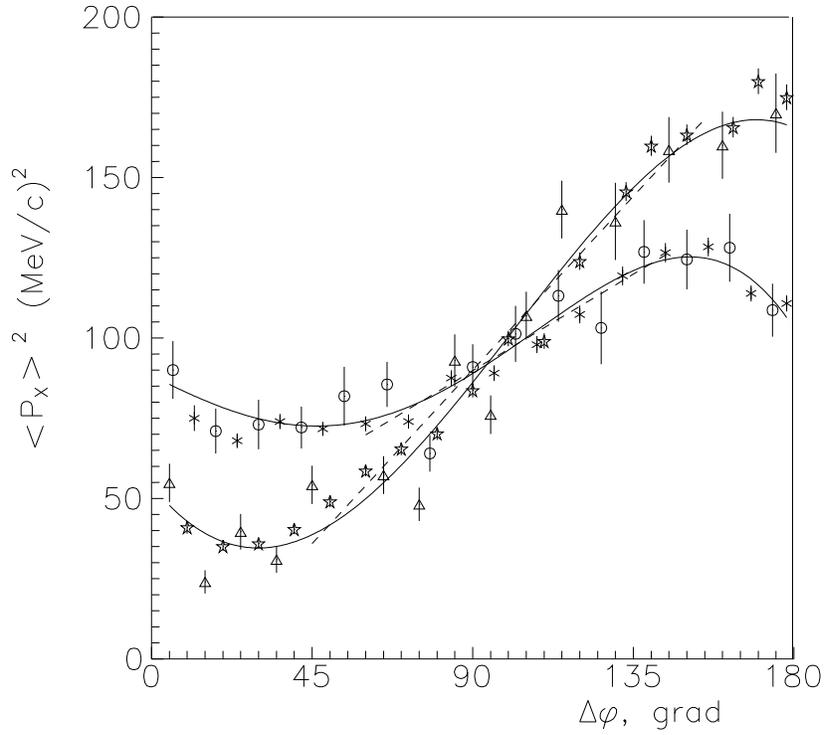,bbllx=0pt,bblly=0pt,bburx=594pt,bbury=842pt,
width=18cm,angle=0}
\end{center}
\vspace{-10.cm}
\hspace{3.cm}
\begin{minipage}{13.0cm}
\caption
{The dependence of $< P_{x} >^{2}$ on the $\Delta \varphi$ 
for protons from: $\circ$ -- C-Ne and $\bigtriangleup$ -- C-Cu
collisions and
{\scriptsize$\times$}, $\star$ -- the QGSM data, respectively.
The dashed lines are the result of the linear approximation of
experimental data in the interval of $\Delta \varphi$ 60 $\div$ 145 
and 45 $\div$ 155,
respectively. The solid curves for visual presentation --
result of approximation by 4-th order polynomial.}
\end{minipage}
\end{figure}
\end{document}